\newcommand{\rmi}[1]{{\mbox{\small #1}}}
\newcommand{\Tr}{\mathop{\textrm{Tr}}}

\newcommand{\re}  {\mathop{\rm Re}}

\documentclass{PoS}

\title{Continuum Study of Heavy Quark Diffusion}

\ShortTitle{Continuum Study of Heavy Quark Diffusion}

\author{\speaker{T. Neuhaus}\thanks{In collaboration with
A.~Francis, O.~Kaczmarek, M.~Laine and H.~Ohno.}\\
        Institute for Advanced Simulation,  
        FZ J\"ulich, D-52425 J\"ulich, Germany\\
        E-mail: \email{t.neuhaus@fz-juelich.de}}

\abstract{We report on a lattice investigation of heavy quark momentum 
diffusion within the pure SU(3) plasma above the deconfinement transition with 
the quarks treated to leading order in the heavy mass expansion. We measure 
the relevant ``colour-electric'' Euclidean correlator and based on several 
lattice spacing's perform the continuum extrapolation. This is necessary 
not only to remove cut-off effects but also the analytic continuation
for the extraction of transport coefficients is well-defined 
only when a continuous function of the Euclidean time variable is available.
We pay specific attention to scale setting in SU(3). In particular we 
present our determination for the critical temperature 
$T_c=1/({N_\tau}a) $ at values of $N_\tau \le 22$.}

\FullConference{9th International Workshop on Critical Point and Onset of Deconfinement - CPOD2014,\\
		17-21 November 2014\\
		ZiF (Center of Interdisciplinary Research), University of Bielefeld, Germany}

\begin{document}

\section{Introduction}

Among the most important quantities playing a role in the theoretical
interpretation of current heavy ion collision experiments at RHIC
(Brookhaven National Laboratory, USA) and LHC (CERN, Switzerland)
are so-called transport coefficients: shear and bulk viscosity's as well
as heavy and light quark diffusion coefficients. Because of strong
interactions, these quantities need to be determined by
non-perturbative numerical lattice Monte Carlo simulations.
This task is a hard one, given that numerical simulations
are carried out in Euclidean signature, whereas transport coefficients
are Minkowskian quantities, necessitating an analytic continuation~\cite{Harvey_Meyer_2011}.
Nevertheless, the problem is solvable in principle~\cite{Cuniberti_2001}, provided that 
lattice simulations reach a continuum limit and that short-distance 
singularities can be subtracted~\cite{Mikko_2011}. We 
present selected results of transport coefficient calculations and 
scale setting in pure $SU(3)$ gauge theory. Scale setting is involved in
in the proper continuum extrapolation of lattice correlation functions.

Most of the results concerning scales and diffusion in the $SU(3)$ plasma
have been published in~\cite{Lattice_2013,Olaf_2014} or will be 
presented elsewhere ~\cite{6,7} . Here we specifically supply additional 
insight in the determination of $T_c$, which at finite temperature is the most fundamental physical 
observable. We leave out our groups high precision studies of the technical 
lattice scales $r_0/a$, the Sommer scale \cite{Sommer_scale_1994} as well as the Wilson flow 
scale $\sqrt{t_0}/a$ \cite{Luescher_scale_2010}. We will however 
make use of the data in \cite{Luescher_scale_2011} and determine the "amplitude product" 
$T_c\sqrt{t_0}$ using $T_c$ data and the L\"uscher et. al. $t_0$ data set.

\section{Colour electric Correlation Function}

Heavy quarks carry a colour charge and, whenever there are gauge 
fields present, are therefore subject to a coloured Lorentz force. 
Like with other transport coefficients the corresponding ``low-energy
constants'' are easiest to define at vanishing three-momentum; then
the Lorentz-force is proportional to the electric field strength. 
This leads to a ``colour-electric correlator''~\cite{Laine_2009,Teaney_2006},  
\begin{equation}
 G_\rmi{\,E}(\tau) = - \frac{1}{3} \sum_{i=1}^3 
 \frac{
  \Bigl\langle
   \re\Tr \Bigl[
      U( \frac{1}{T} ,\tau) \, gE_i(\tau,\vec{0}) \, U(\tau,0) \, gE_i(0,\vec{0})
   \Bigr] 
  \Bigr\rangle
 }{
 \Bigl\langle
   \re\Tr [U( \frac{1}{T} ,0)] 
 \Bigr\rangle
 \label{propagator}
 }
\end{equation}
where $gE_i$ denotes the colour-electric field, $T$ the temperature, and $U(\tau_2,\tau_1)$
a Wilson line in Euclidean time direction. If the corresponding spectral function, $\rho_\rmi{\,E}$,  
can be extracted~\cite{Harvey_Meyer_2011}, then the ``momentum diffusion coefficient'', often 
denoted by $\kappa$, can be obtained from 
\begin{equation}
 \kappa = \lim_{\omega\to 0} \frac{2 T \rho_\rmi{\,E}(\omega)}{\omega}
 \;. 
\end{equation}
According to non-relativistic linear response relations (valid 
for $M \gg \pi T$, where $M$ stands for a heavy quark pole mass)
the corresponding ``diffusion coefficient'' is given by $D = 2 T^2/\kappa$. We determine
stochastic estimates for the function $G_\rmi{\,E}(\tau)$ from large scale Monte Carlo 
simulations in finite temperature $T \approx 1.45 T_c$ pure gauge $SU(3)$ lattice gauge 
theory with standard Wilson action and for box sizes, that substantially exceed 
those of earlier simulation results \cite{Olaf_2011}. 
The largest volume is $V_{\rm max}=48 \times 192^3$ which is simulated with 
sufficient statistics at $\beta=7.793$. Table~1 shows run parameters with  $N_{\mathrm{stat}}$
labeling the number of statistically independent configurations, while  $r_0 T$
denotes an estimate of the Sommer scale in units of the temporal box extent. 
\begin{table}
\centering{
\begin{tabular}{|c|c|c|c|c|} 
\hline      
$\beta$ & $N_\tau$ & $N_s$ & $N_{\mathrm{conf}}$ & $r_0 T$  \\
\hline\hline
     6.872 & 16 & ~64 & 100 & 1.111 \\
     7.192 & 24 & ~96 & 160 & 1.077 \\
     7.544 & 36 & 144 & 563 & 1.068 \\
     7.793 & 48 & 192 & 223 & 1.055 \\
\hline
\end{tabular}}
\caption{A lattice discretized version of the correlator eq.~2.1
is calculated at the given $\beta$ values and box sizes for the finite high temperature
$T \approx 1.45 T_c$.}
\end{table}
\begin{figure}
\centerline{\includegraphics[height=0.35\textwidth, width=0.50\textwidth]{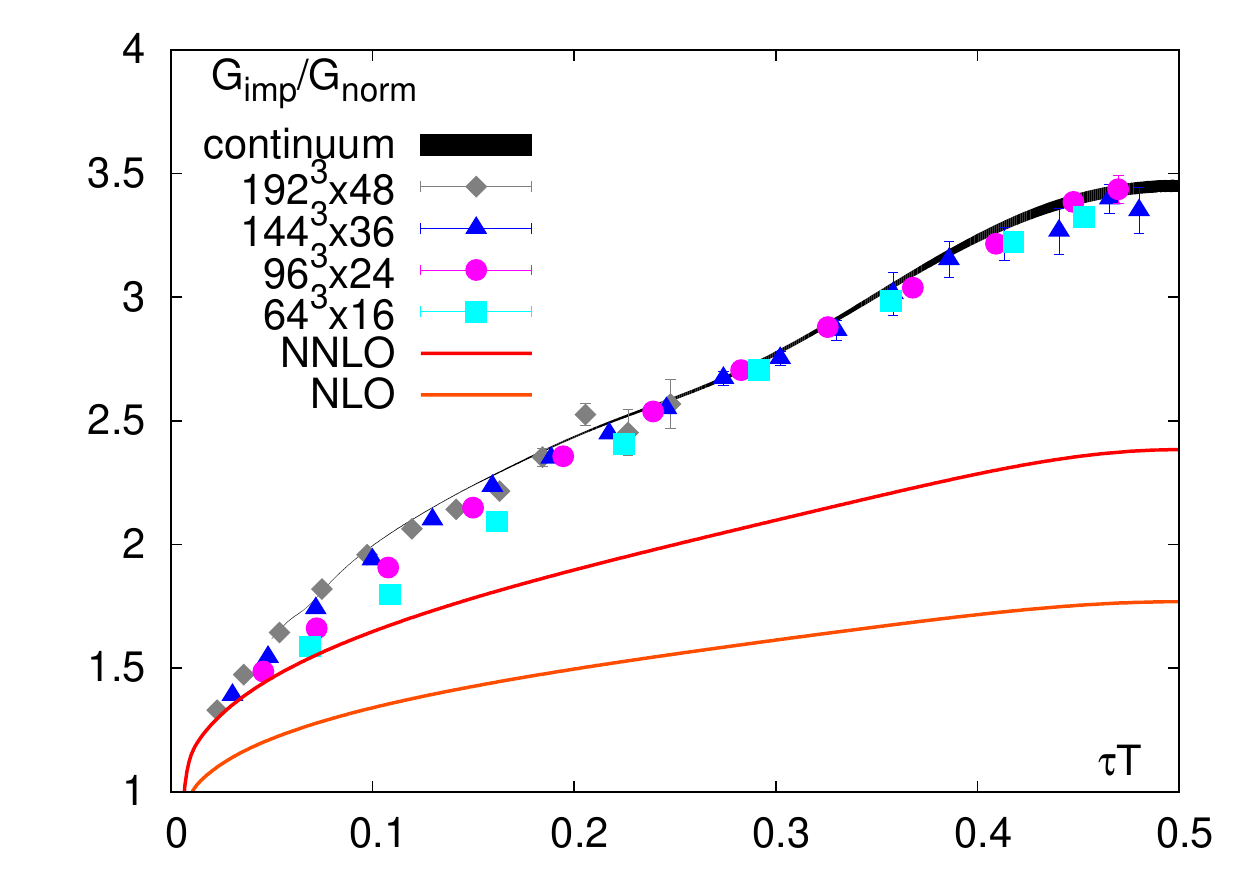}}
\caption{Correlation function ratio $G_{imp}(\tau T)/ G_\rmi{norm}(\tau T)$, see text, with perturbative 
NLO \cite{20} and NLLO calculations. The black band denotes the desired continuum extrapolation result.}
\end{figure}
We perform quenched QCD calculations for the discrete version
of the correlation function $G_E(\tau)$. Due to the bad signal to noise
ratio of the purely gluonic operator and the rapid decrease of the correlation function
with $\tau$ we have to use noise reduction techniques i.e., multi-level updates 
\cite{multi_level_2001,multi_level_2007} and link-integration ``PPR'' \cite{17,18}. 
Furthermore we use a tree-level improvement \cite{19} to reduce cut-off
effects and a NLO perturbative renormalization factor $Z_{pert}(\beta)$. In Fig.~1 the lattice 
results for $G_{imp}(\tau T)$ are shown, normalized to 
\begin{equation}
 G_\rmi{norm}(\tau T)
 \; \equiv \;
 \pi^2 T^4 \left[
 \frac{\cos^2(\pi \tau T)}{\sin^4(\pi \tau T)}
 +\frac{1}{3\sin^2(\pi \tau T)} \right]
 \;.
\end{equation}
Although cut-off effects are visible at small separations
and the results become more noisy at large distances on the finer 
lattices, the results on the four lattices allow for a controlled 
continuum extrapolation down to distances around $\tau T\sim
0.05$. To obtain the continuum estimate of the correlation 
function, we perform b-spline or polynomial interpolations for 
each lattice and at fixed $\tau T$ extrapolate the correlator 
in $1/N_\tau^2$. To allow for needed contributions to the 
spectral function $\rho_{\rm E}(\omega)$ we use the Ansatz
\begin{eqnarray}
\rho_{model}(\omega) = \max\left\{ A \rho_{NNLO}(\omega) + B \omega^3, \frac{\omega\kappa}{2 T}\right\},
\end{eqnarray}
containing three parameters $A$, $B$ and $\kappa$, and fit the Euclidean correlator with the form
\begin{eqnarray} 
 G_{model}(\tau) =
 \int_0^\infty
 \frac{{\rm d}\omega}{\pi} \rho_{model}(\omega)
 \frac{\cosh \left(\frac{1}{2} - \tau T \right)\frac{\omega}{T} }
 {\sinh\frac{\omega}{2 T}} 
 \;
\end{eqnarray} 
in the $\tau T$ range [0.1:0.5] to the continuum. Under the given procedure we 
find the momentum diffusion coefficient:
\begin{eqnarray}
\kappa/T^3 = 2.5(4)~~~T \approx 1.45 T_c.
\end{eqnarray}
For a more detailed presentation we refer here to \cite{Olaf_2014}. We will
have to investigate the stability of the numeric value under
various model assumptions for the spectral density.   

\section{Deconfinement Position at large temporal Extent }

Using alternate finite size scaling arguments \cite{21} from 
statistical physics, based on phase weight ratios at first order 
phase transitions we determine the finite temperature transition 
couplings $\beta_c(N_{\tau})$ for a maximal value $N_\tau=22$. The
numerical findings are contained in Table~2. 
\begin{table}
\centerline{
\begin{tabular}{|c|c|}
\hline
$N_\tau$ & $\beta_c$ \\
\hline
4   &  5.69275(28) \\
6   &  5.89425(29) \\
8   &  6.06239(38) \\
10  &  6.20872(47) \\
12  &  6.33514(45) \\
14  &  6.4473(18)  \\
16  &  6.5457(40)  \\
18  &  6.6331(20)  \\
20  &  6.7132(26)  \\
22  &  6.7986(65)  \\
\hline
\end{tabular}}
\label{the_table}
\caption{Numerical $\beta_c$ values for pure gauge $SU(3)$ theory with Wilson action.}
\end{table}
\begin{figure}
\vspace{-1.00cm}
\centerline{\includegraphics[height=0.50\textwidth, width=0.50\textwidth]{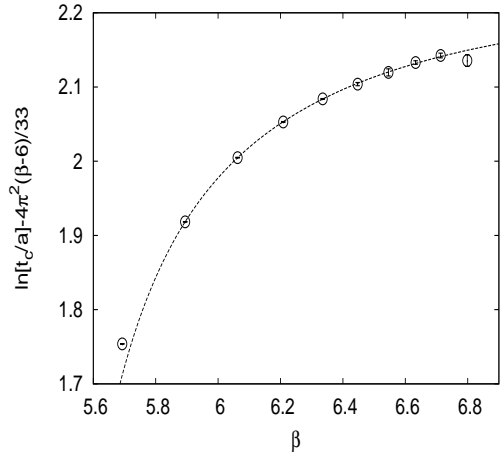}}
\vspace{-1.00cm}
\caption{Given the critical time extent $t_c/a=N_\tau$ 
as calculated from Table~2 we display the 
quantity ${\rm ln}(t_c/a) -4\pi^2(\beta-6)/33$ as a function of $\beta$. 
The curve for $\beta>5.8$ ($N_\tau > 4$) displays the fitted rational Ansatz eq.~3.1, see the text.}
\end{figure}
The $\beta_c$ data of Table~2 correspond directly to the 
circles in Fig.~2 where the temporal time extent $t_c/a=N_\tau$
in units of $a$ at the $SU(3)$ Deconfinement phase transition is suitably logarithmized
and subtracted: ${\rm ln}(t_c/a) -4\pi^2(\beta-6)/33$, and then is 
displayed as a function of $\beta$. Error propagation via 
$\delta t_c \approx \delta\beta_cN_\tau(4\pi^2/33)$
was used. It has become customary, as in case for the Sommer scale $r_0$, to 
interpolate logarithmized QCD scales ${\rm ln}(s/a)$ with a rational Ansatz \cite{rational_interpolation_2007}
\begin{equation}
{\rm ln}({s \over a})=\lbrack{\beta \over 12b_0}+{b_1 \over 2b_0^2}{\rm ln}({6b_0 \over \beta})\rbrack ({ 1+c_1\beta^{-1}+c_2\beta^{-2}\over  1+c_3\beta^{-1}+c_4\beta^{-2}})
\end{equation}
at values $b_0=11/(4\pi)^2$ and $b_1=102/(4\pi)^4$.
We find the fit parameter values 
\begin{equation}
c_1=-9.5347(2060)~~~c_2=22.252(1.053)~~~c_3=-6.5581(2958)~~~c_4=6.547(1.241) 
\end{equation}
at a reduced $\chi^2_{\rm d.o.f} =0.87$ for the fit if all data points with $N_\tau \ge 6$ are included.
See the curve in Fig.~2.

The finite size method for $\beta_c$ splits the phase space of the 
complex Polyakov loop into an outer region (Deconfinement phase)
and an inner (Confined phase) and determines the coupling $\beta$
at which point the weight of the phases is $3$ to $1$, because of
$SU(3)$'s $Z(3)$ center symmetry. A proper defined step function of weights 
$w_c$ and $w_d$ with $s(\beta)=(3w_c-w_d)/(3w_c+w_d)$
see  Fig.~3, then has a zero which is determined by a linear fit.
Our finite size determinations are all done with boxes $N_s > 3 N_\tau$
except for $N_\tau=22$ at $N_s=64$. For values $N_\tau \le 16$ we 
have checked, that the weight ratio method results are consistent with polynomial 
finite size scaling extrapolations of susceptibility peak positions 
at the first order phase transition. The weight ratio method 
however, and already for $N_s \ge 3 N_\tau$,
has smaller corrections, which only are exponentially small
in the spacial size. Our $\beta_c$ data in Table~2 exert stress on semi-analytic 
calculations \cite{semi_analytic_Philipsen,semi_analytic_Tomboulis}
but are consistent with the Bielefeld numbers 
\cite{bielefeld_critical_points_1996,bielefeld_critical_points_1999}
as well as with Berg et. al. \cite{berg_critical_points_2013}. Deviations
mostly lie within the $1~\sigma$ error margin.
\begin{figure}
\vspace{-1.00cm}
\centering{\includegraphics[height=0.5\textwidth, width=0.5\textwidth]{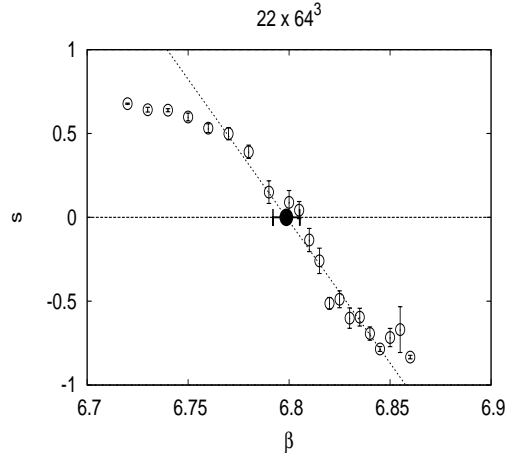}}
\vspace{-1.00cm}
\caption{For the $N_\tau=22$ and $N_s=64$ box we display as a function of $\beta$
the step function $s(\beta)$, whose zero determines the critical point 
at $N_\tau=22$ as in Table~2. The zero is marked by a solid circle and horizontal error bar 
and is obtained by the straight line fit, the line in the figure.}
\end{figure}

In finite temperature lattice calculations, it is 
wishful to present the critical temperature $T_{\rm c}$ 
in physical units. If there are no precise 
measurements of physical observables like
e.g. the string tension we can also
opt for less physical zero temperature lattice scales. We determine
two scales: the Sommer scale \cite{Sommer_scale_1994} denoted by $r_0/a$
and the Wilson flow $\sqrt{t_0}/a$ as recently introduced by L\"uscher 
\cite{Luescher_scale_2010,Luescher_scale_2011} and leave the presentation
to forthcoming work \cite{6}.
In case of $t_0$ and for any given $\beta$, the lattice configuration
is ``cooled'' using Runge Kutta algorithm for a classical Wilson flow 
until a stopping criterion on the energy density is reached at 
pseudo time $t_0$. Various different regularization's 
(Wilson, Wilson improved, Clover) for the energy density can be employed.
Here we will use L\"uscher's $t_0$ numbers 
from \cite{Luescher_scale_2011} for the range $5.96 \le \beta \le 6.59$, corresponding to 
the $N_\tau$ range $8 \le N_\tau \le 18$. We interpolate ${\rm ln}(\sqrt{t_0}/a)$ similar to 
${\rm ln}(s/a)$ as in eq.~3.1 by a rational form and determine $T_c\sqrt{t_0}$
in the continuum limit $a^2/t_0 \rightarrow 0$. 
\begin{figure}
\vspace{-1.00cm}
\centering{\includegraphics[height=0.5\textwidth, width=0.5\textwidth]{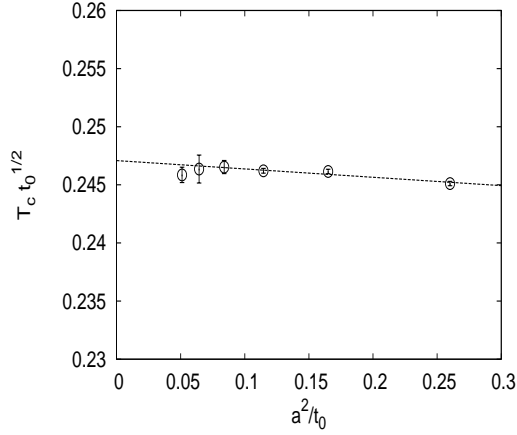}}
\vspace{-1.00cm}
\caption{$T_c\sqrt{t_0}$ calculated from data in Table~2 at $N_\tau=8,...,18$ (data points
from the right to the left). The $t_0$ data of \cite{Luescher_scale_2011} are used to interpolate
$t_0$ by the rational form of eq.~3.1.}
\end{figure}
We display in Fig.~4 the continuum extrapolation for $T_c\sqrt{t_0}$ based on the L\"uscher data.
A straight line fit, the line in Fig.~4, through 
all six data points with order ${\cal O}(a^2)$ corrections only,  yields:
\begin{eqnarray}
T_c\sqrt{t_0}~~~=~~~0.2471(3)[10]~~~~~\chi^2_{d.o.f}=0.94.
\end{eqnarray}
We observe very small $a^2$ corrections of relative magnitude $0.004$ 
accompanied by a small statistical error $0.0003$. Varying the fit range, and 
also adding $a^4$ corrections to the fit, we estimate an systematic uncertainty
of $0.0010$ in the observable and thus $T_c\sqrt{t_0}=0.2471(13)$ is a realistic 
finding. It is to be remarked, that the scaling deviations in the amplitude product 
turn out to be quite smooth and regular. Thus, under the provision of 
correct and precise $t_0$ data \footnote{The values of $t_0/a^2$ in \cite{Luescher_scale_2011} 
are probably affected by downward finite size effects of the order few percent, as 
the authors concede.}, our $T_c$ determinations for $N_\tau \le 18$
appear to be of similar numerical quality. Naturally the physical 
temperature is more relevant than $t_0$, unless $t_0$ can be converted 
into another physical scale at same precision levels.

\section{Conclusion}
We have estimated the critical couplings $\beta_c$ for the Wilson action and 
for temporal extents up to $N_\tau=20$ with relative precision levels  
$\le 0.001$. For the $N_\tau$ interval $6,...,22$
we also provide a rational interpolation to ${\rm ln}(t_c/a)$
in eq.~3.2 and eq.~3.3, on the same footing as for the Sommer scale in  \cite{rational_interpolation_2007}, 
$t_c$ being the critical time extent. It remains to be seen 
how the determination of the diffusion constant $\kappa$
in eq.~2.6 responds to more elaborate models for the spectral function and 
depends on the finite quark mass and on the continuum limit. 
Work in this direction is in progress

{\bf Acknowledgements:} Simulations where performed on the JARA0039 and
JARA0108 accounts of JARA-HPC in Aachen, JUDGE/JUROPA at JSC J\"ulich, 
the OCuLUS Cluster in Paderborn, and the Bielefeld GPU cluster.

\end{document}